\documentclass[twocolumn]{aastex62}
\pdfoutput=1 
\usepackage{amsmath,amstext}
\usepackage[T1]{fontenc}
\usepackage[figure,figure*]{hypcap}
\usepackage[version=4]{mhchem} 

\newcommand{\rosetta}{{\sl Rosetta}}

\shorttitle{\rosetta\ Stellar Occultation}
\shortauthors{Keeney et~al.}

\turnoffedits
\submitjournal{\aj}
\received{2018 November 8}
\revised{2019 February 25}
\accepted{2019 March 15}

\begin{document}

\title{Stellar Occultation by Comet 67P/Churyumov-Gerasimenko Observed with \\ \rosetta's Alice Far-Ultraviolet Spectrograph}

\author{Brian A. Keeney}
\affiliation{Southwest Research Institute, Department of Space Studies, Suite 300, 1050 Walnut St., Boulder, CO 80302, USA; bkeeney@gmail.com}
\author{S. Alan Stern}
\affiliation{Southwest Research Institute, Department of Space Studies, Suite 300, 1050 Walnut St., Boulder, CO 80302, USA; bkeeney@gmail.com}
\author{Paul D. Feldman}
\affiliation{Department of Physics and Astronomy, Johns Hopkins University, 3400 N. Charles St., Baltimore, MD 21218, USA}
\author{Michael F. A'Hearn}
\altaffiliation{Deceased}
\affiliation{Department of Astronomy, University of Maryland, College Park, MD 20742, USA}
\author{Jean-Loup Bertaux}
\affiliation{LATMOS, CNRS/UVSQ/IPSL, 11 Boulevard d'Alembert, F-78280 Guyancourt, France}
\author{Lori M. Feaga}
\affiliation{Department of Astronomy, University of Maryland, College Park, MD 20742, USA}
\author{Matthew M. Knight}
\affiliation{Department of Astronomy, University of Maryland, College Park, MD 20742, USA}
\author{Richard A. Medina}
\affiliation{Southwest Research Institute, Department of Space Studies, Suite 300, 1050 Walnut St., Boulder, CO 80302, USA; bkeeney@gmail.com}
\author{John Noonan}
\affiliation{Lunar and Planetary Laboratory, University of Arizona, 1629 E. University Blvd., Tucson, AZ 85721, USA}
\author{Joel Wm. Parker}
\affiliation{Southwest Research Institute, Department of Space Studies, Suite 300, 1050 Walnut St., Boulder, CO 80302, USA; bkeeney@gmail.com}
\author{Jon P. Pineau}
\affiliation{Stellar Solutions, Inc., 250 Cambridge Ave., Suite 204, Palo Alto, CA 94306, USA}
\author{Rebecca N. Schindhelm}
\affiliation{Southwest Research Institute, Department of Space Studies, Suite 300, 1050 Walnut St., Boulder, CO 80302, USA; bkeeney@gmail.com}
\affiliation{Ball Aerospace and Technology Corp., 1600 Commerce St., Boulder, CO 80301, USA}
\author{Andrew J. Steffl}
\affiliation{Southwest Research Institute, Department of Space Studies, Suite 300, 1050 Walnut St., Boulder, CO 80302, USA; bkeeney@gmail.com}
\author{M. Versteeg}
\affiliation{Southwest Research Institute, 6220 Culebra Rd., San Antonio, TX 78238, USA}
\author{Ronald J. Vervack, Jr.}
\affiliation{Space Exploration Sector, Johns Hopkins University Applied Physics Laboratory, 11100 Johns Hopkins Rd., Laurel, MD 20723, USA}
\author{Harold A. Weaver}
\affiliation{Space Exploration Sector, Johns Hopkins University Applied Physics Laboratory, 11100 Johns Hopkins Rd., Laurel, MD 20723, USA}

\begin{abstract}
Following our previous detection of ubiquitous \ce{H2O} and \ce{O2} absorption against the far-UV continuum of stars located near the nucleus of Comet~67P/Churyumov-Gerasimenko, we present a serendipitously observed stellar occultation that occurred on 2015 September~13, approximately one month after the comet's perihelion passage. The occultation appears in two consecutive 10-minute spectral images obtained by Alice, \rosetta's ultraviolet (700-2100~\AA) spectrograph, both of which show \ce{H2O} absorption with column density $>10^{17.5}~\mathrm{cm}^{-2}$ and significant \ce{O2} absorption ($\ce{O2}/\ce{H2O} \approx 5$-10\%). Because the projected distance from the star to the nucleus changes between exposures, our ability to study the \ce{H2O} column density profile near the nucleus (impact parameters $<1$~km) is unmatched by our previous observations. We find that the \ce{H2O} and \ce{O2} column densities decrease with increasing impact parameter, in accordance with expectations, but the \ce{O2} column decreases $\sim3$ times more quickly than \ce{H2O}. When combined with previously published results from stellar appulses, we conclude that the \ce{O2} and \ce{H2O} column densities are highly correlated, and $\ce{O2}/\ce{H2O}$ decreases with increasing \ce{H2O} column.
\end{abstract}


\section{Introduction}
\label{sec:intro}

The double-focusing mass spectrometer (DFMS) of the \rosetta\ Orbiter Spectrometer for Ion and Neutral Analysis \citep[ROSINA;][]{balsiger07} has found that \ce{O2} is the fourth most abundant parent species in the coma of Comet~67P/Churyumov-Gerasimenko (67P/C-G), behind only \ce{H2O}, \ce{CO2}, and \ce{CO} \citep{leroy15,fougere16}. The ubiquitous, abundant presence of \ce{O2} was surprising since it had never been detected in a comet before \citep{bieler15}.

\citet{feldman16} confirmed the presence of substantial \ce{O2} in the coma of 67P/C-G during gaseous outbursts with Alice, \rosetta's ultraviolet spectrograph \citep{stern07}. Later, \citet{keeney17} directly detected \ce{O2} absorption in Alice data using stellar sight lines temporarily projected near the nucleus (``stellar appulses''). These sight lines were observed over a wide range of heliocentric distances (1.2-2.3~AU) and impact parameters (4-20~km), yielding $\log{N_{\ce{H2O}}}=15.2$-17.1 (all column densities, $N$, herein are quoted in units of $\mathrm{cm}^{-2}$) and a median value of $N_{\ce{O2}}/N_{\ce{H2O}}=25$\%.

Several \rosetta\ instruments can directly detect \ce{H2O} in the coma of 67P/C-G: Alice, ROSINA, the Visible and Infrared Thermal Imaging Spectrometer \citep[VIRTIS;][]{coradini07}, and the Microwave Instrument for the \rosetta\ Orbiter \citep[MIRO;][]{gulkis07}. All but ROSINA are remote-sensing instruments that measure or infer column densities along a line of sight, and $N_{\ce{H2O}}$ measured by Alice \citep{keeney17} agrees with VIRTIS values near perihelion \citep{bockelee-morvan16}. However, only Alice and ROSINA can directly detect \ce{O2}, and the relative abundance of $\ce{O2}/\ce{H2O}$ in the Alice data was nearly an order of magnitude larger than the average ROSINA value \citep[$n_{\ce{O2}}/n_{\ce{H2O}}=3.85\pm0.85$\%;][where $n$ is number density measured at the spacecraft position]{bieler15}. The consistency in $N_{\ce{H2O}}$ between Alice and VIRTIS suggests that the methodology of \citet{keeney17} is trustworthy, but the discrepancy in $\ce{O2}/\ce{H2O}$ remains puzzling.

\begin{figure}
  \epsscale{1.15}
  \centering\plotone{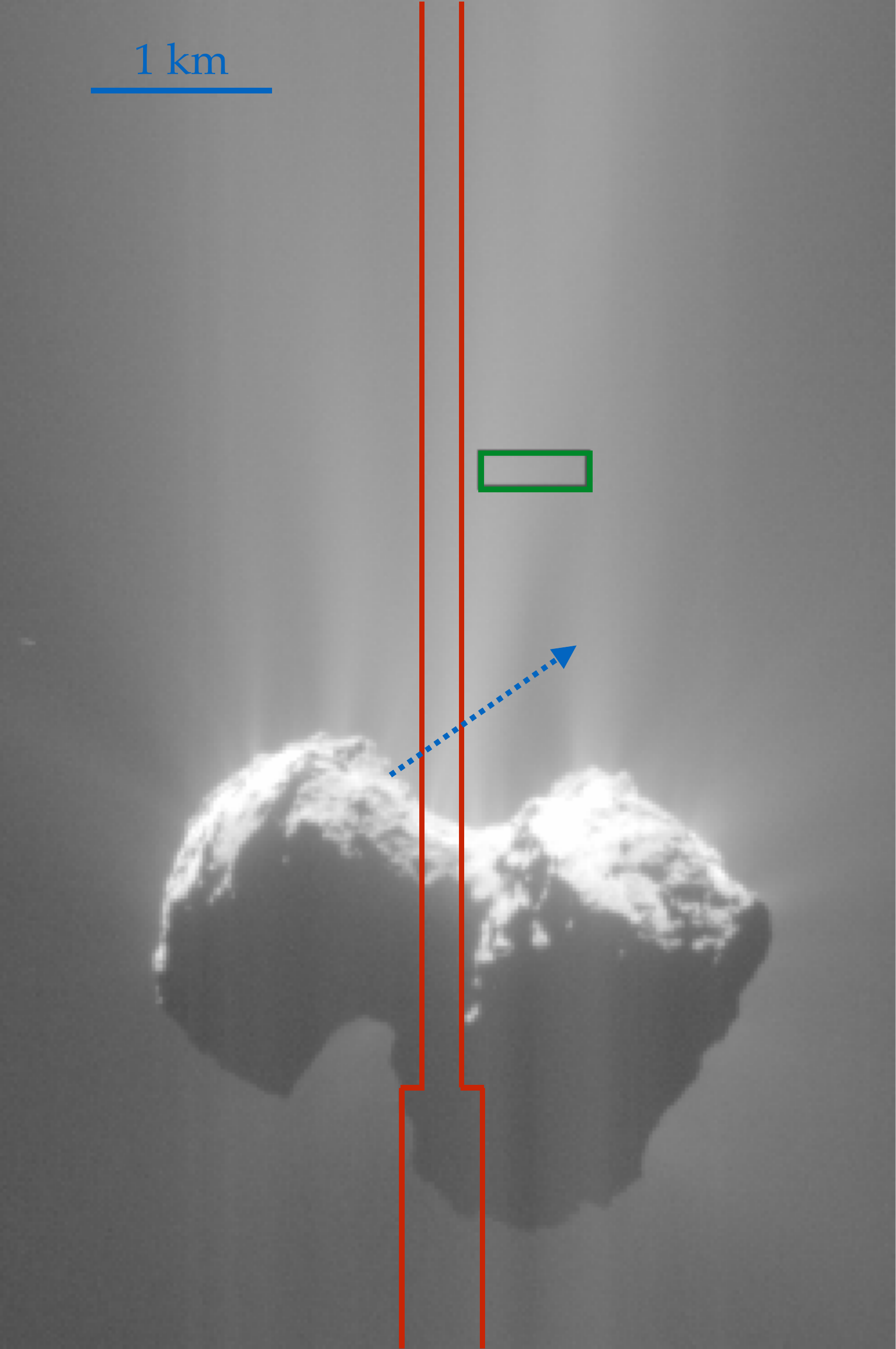}
  \vspace{-1ex}
  \caption{Subsection of a NAVCAM image taken while HD~4150 was occulted by 67P/C-G, $\sim20$~minutes before the star entered the Alice slit (red). The VIRTIS-H aperture is shown in green, and the approximate path of HD~4150 through the Alice slit is shown by a dashed blue arrow. The image is oriented such that the Sun is toward the top, and the field-of-view is $\sim1\degr\times1\fdg5$ ($\sim5\times7.5~\mathrm{km}^2$).
  \label{fig:navcam}}
\end{figure}

Here we present a stellar occultation by 67P/C-G observed with Alice. We describe our observations in \autoref{sec:obs}, and our analysis procedure in \autoref{sec:results}. \autoref{sec:disc} compares the \ce{H2O} and \ce{O2} column densities for our stellar occultation with those of \citet{keeney17} and ROSINA measurements \citep{bieler15,hansen16}, and \autoref{sec:summary} summarizes our main findings.

\section{Observations}
\label{sec:obs}

The A0~IV star HD~4150 was occulted by 67P/C-G on 2015 September~13, approximately one month after the comet's perihelion passage. The occultation was serendipitously observed by Alice during the course of normal operations. During this phase of the mission, Alice was integrating nearly continuously to catalog emissions from the near-nucleus coma \citep{pineau19}.

\autoref{fig:navcam} shows a $\sim5\times7.5~\mathrm{km}^2$ subsection of a \rosetta\ navigation camera (NAVCAM) image taken at 13:36:02~UT, while HD~4150 was occulted by 67P/C-G and $\sim20$~minutes before it entered the Alice slit. The Alice slit is $5\fdg5$ long and has a dog-bone shape that is twice as wide at the top and bottom as in the center \citep{stern07}; its position with respect to the nucleus is shown in red in \autoref{fig:navcam}, and the transition between the wide-bottom and narrow-center regions of the slit is evident. \autoref{fig:navcam} also shows the approximate path of HD~4150 as it emerges from behind the small lobe of 67P/C-G and crosses the Alice slit from left to right at an angle of $\sim35\degr$ over the ``neck'' of the nucleus.

\begin{figure}
  \epsscale{1.15}
  \centering\plotone{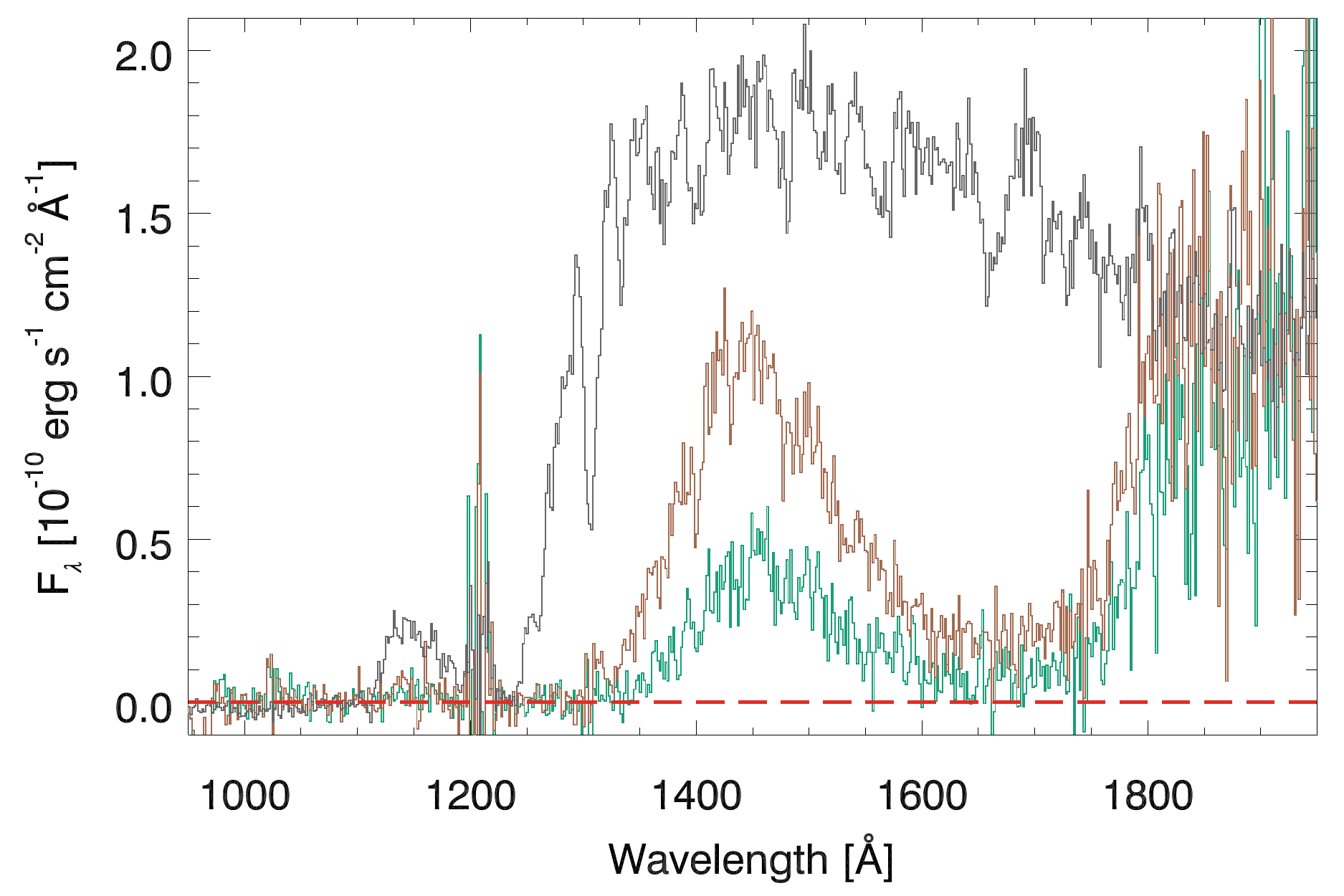}
  \vspace{-1ex}
  \caption{Three spectra of HD~4150 obtained by Alice. The revisit (i.e., intrinsic) spectrum \edit1{taken on 2016 June~6} is shown in gray, and the \edit1{post-occultation spectra taken on 2015 September~13 at 13:52:04 and 14:02:49~UT} are shown in green and brown, respectively.
  \label{fig:exp}}
\end{figure}

HD~4150 appears in two consecutive 10-minute spectral images with start times of 13:52:04 and 14:02:49~UT, respectively. For the first five minutes of the first exposure, the star is occulted by the comet nucleus. The star remains in the Alice slit for the remainder of the first exposure and the first five minutes of the second exposure. During these exposures, 67P/C-G was 1.30~AU from the Sun, \rosetta\ was orbiting 313~km from the comet center, and the solar phase angle was $108\degr$.

The spectra of HD~4150 extracted from these exposures are shown in green (first exposure) and brown (second exposure) in \autoref{fig:exp}. The observed stellar fluxes are corrected for the reduced amount of time the star was in the slit during these exposures. Nevertheless, differences between the two exposures are evident; most notably, the first (green) spectrum has less flux from 1350-1600~\AA.

HD~4150 was re-observed (``revisited'') on 2016 June~6 when it was far from the nucleus, at an off-nadir angle of $79\degr$. The purpose of this 15-minute integration, which started at 06:02:24~UT, was to characterize the intrinsic stellar spectrum without the presence of foreground coma absorption (see \citealp{keeney17} for details). When HD~4150 was revisited, the comet was 3.15~AU from the Sun and the solar phase angle was $67\degr$. The intrinsic stellar spectrum is shown in gray in \autoref{fig:exp}, and has considerably more flux from 1250-1800~\AA\ than the spectra obtained immediately after the occultation.

The analysis below, and that of \citet{keeney17}, assumes that the stellar spectrum obtained when the star was revisited while far from the nucleus is equivalent to the intrinsic stellar spectrum (i.e., that there is no foreground coma absorption at that time). Although it is true that \ce{O2} is extremely volatile, with a sublimation temperature in vacuum of $\sim30$~K \citep{fray09}, it has been found to be strongly correlated with \ce{H2O} in the coma of 67P/C-G \citep{bieler15,fougere16}, and the production rate of \ce{H2O} was down by 2-3 orders of magnitude at 3.15~AU compared to 1.30~AU \citep{hansen16,gasc17}. However, even if our assumptions are incorrect and a small amount of foreground coma absorption is present in the revisit spectrum, then this would cause us to \textit{underestimate} the amount of foreground coma absorption at the time of the occultation\footnote{The same can be said for the stellar appulse analysis of \citet{keeney17}.}.

\section{Results and Analysis}
\label{sec:results}

The spectra of HD~4150 taken immediately after the occultation by 67P/C-G were analyzed analagously to the stellar appulse spectra of \citet{keeney17}, except in one regard. The UV-bright stars that \citet{keeney17} used to study the near-nucleus coma all had measurable far-UV flux down to $\sim900$~\AA\ \citep[see Figure~1 of][]{keeney17}, but HD~4150 is an A0 star with almost no flux below $\sim1250$~\AA\ (see \autoref{fig:exp}). This lack of flux blueward of 1250~\AA\ has important consequences for our analysis.

Although \ce{H2O} and \ce{O2} have relatively large cross sections from 1250-1800~\AA\ \citep{chung01,yoshino05}, other abundant coma species do not \citep[i.e., \ce{CO} and \ce{CO2};][]{cairns65,yoshino96}, which means that they cannot be directly constrained by our data. Fortunately, at the time of the occultation VIRTIS was acquiring data while pointed $\sim3.6$~km above the sunward limb of the nucleus. Thus, we adopt their contemporaneous column density ratio of $N_{\ce{CO2}}/N_{\ce{H2O}} = 0.310\pm0.034$ \citep{bockelee-morvan16} for our analysis. The position of the VIRTIS-H aperture with respect to the Alice slit and the path of HD~4150 are shown in \autoref{fig:navcam}.

To isolate the coma signature from the intrinsic stellar flux and interstellar absorption, we divided the stellar spectra taken immediately post-occultation (i.e., the green and brown spectra in \autoref{fig:exp}) by the spectrum of the star taken much later (i.e., the gray spectrum in \autoref{fig:exp}), after first scaling them to have the same median flux from 1850-1950~\AA. This procedure reduces our sensitivity to the uncertainty in the amount of time the star was in the slit during our long exposures \citep{keeney17}. The resulting normalized spectrum quantifies the amount of foreground coma absorption in the post-occultation exposures.

\citet{keeney17} modeled far-UV absorption from ten molecular species (see their Table~3 and Figure~2 for adopted cross sections) in Alice spectra normalized as above. We use the same procedure here, except that we only fit wavelengths in the range 1250-1950~\AA\ because HD~4150 has insufficient flux blueward of 1250~\AA\ (see \autoref{fig:exp}). Consequently, we remove \ce{CO} from our fits because it has no appreciable cross section redward of 1050~\AA\ \citep{cairns65}.

The column density of \ce{H2O} is fit directly and allowed to vary in the range $\log{N_{\ce{H2O}}} = 14$-18. The abundances of all other species are fit relative to $N_{\ce{H2O}}$; \ce{O2} is allowed to vary in the range $N_{\ce{O2}}/N_{\ce{H2O}}=0$-1, \ce{CO2} is fixed at the ratio measured by VIRTIS \citep[$N_{\ce{CO2}}/N_{\ce{H2O}} = 0.310\pm0.034$;][]{bockelee-morvan16}, and all other species (\ce{CH4}, \ce{C2H2}, \ce{C2H6}, \ce{C2H4}, \ce{C4H2}, and \ce{H2CO}) are constrained to have $N/N_{\ce{H2O}}<0.01$ \citep{leroy15}. All species except \ce{CO} and \ce{CO2} are treated the same as they were in \citet{keeney17}. \edit1{The absorption profiles are determined directly from high-resolution cross sections \citep[1-2~\AA\ for \ce{H2O} and \ce{O2}; references for all adopted cross sections are listed in Table~3 of][]{keeney17}, then convolved to the spectral resolution of Alice (9~\AA\ FWHM for the narrow part of the slit) before being compared to the data.}

\begin{figure}
  \epsscale{1.15}
  \centering\plotone{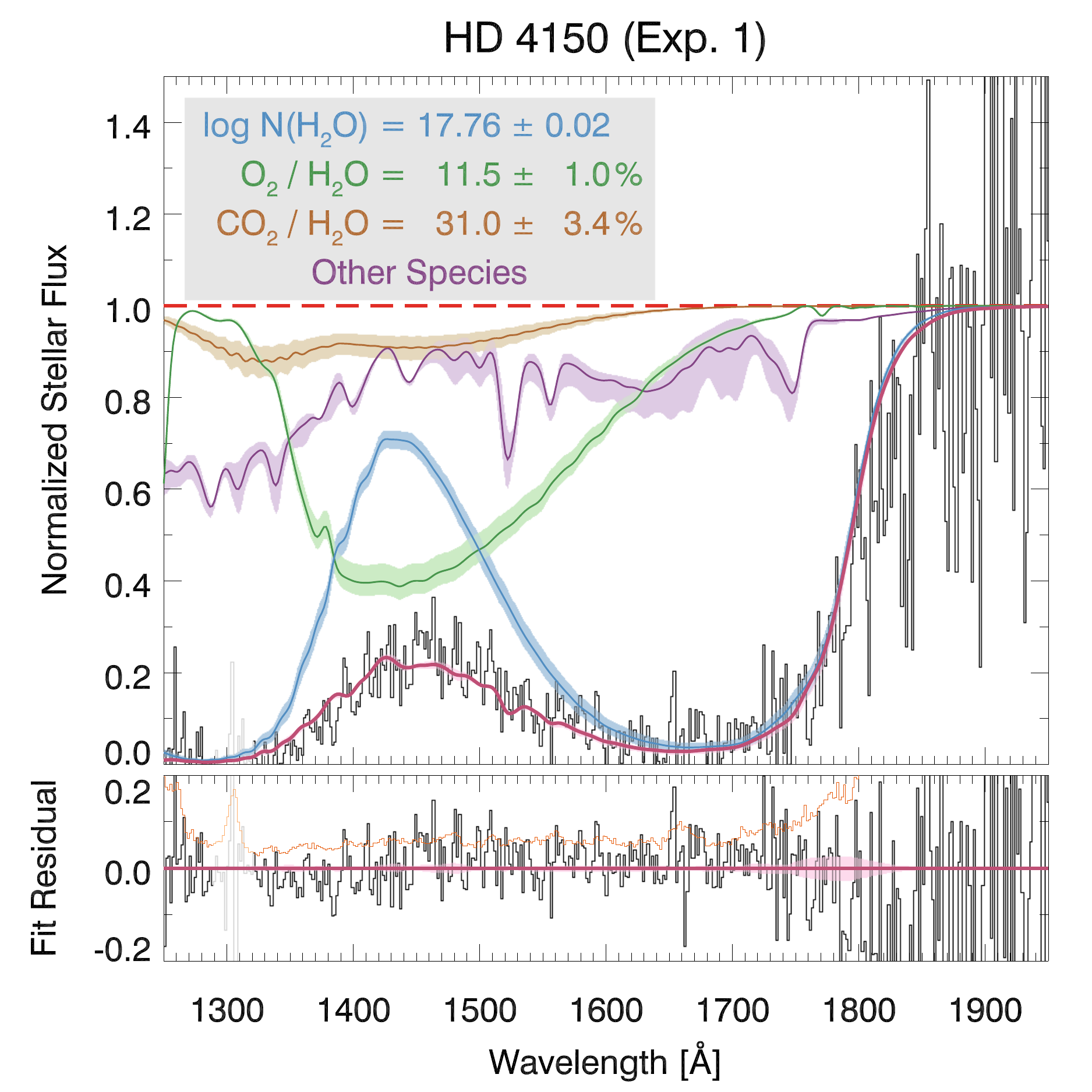}
  \vspace{-1ex}
  \caption{Best-fit column densities for the first post-occultation exposure of HD~4150, with 95\% ($2\sigma$) confidence bands. \textit{Top:} the normalized stellar flux with ensemble fit (pink) and individual-species absorption overlaid. \textit{Bottom:} the residual of the ensemble fit with $1\sigma$ flux uncertainty (orange) overlaid. Masked regions are shown in lighter hues in both panels; these regions are not used to constrain the fits.
  \label{fig:fit1}}
\end{figure}

Our fits to the spectra of HD~4150 taken immediately after its occultation by 67P/C-G are shown in \autoref{fig:fit1}-\ref{fig:fit2}. The top panels show the normalized spectra in black, and our best-fit absorption profiles from \ce{H2O} (blue), \ce{O2} (green), \ce{CO2} (brown), and all other species (purple). The ensemble fit from all species is shown in pink, and the shaded regions represent 95\% ($2\sigma$) confidence bands. Recall that $N_{\ce{CO2}}/N_{\ce{H2O}}$ is held fixed at the value measured contemporaneously by VIRTIS \citep{bockelee-morvan16}. The bottom panel shows the residual of the ensemble fit as a function of wavelength, with $1\sigma$ flux uncertainty overlaid in orange. Masked regions that are not used to constrain the fits are shown in lighter hues in both panels.

We searched for systematic offsets in the best-fit \ce{H2O} and \ce{O2} column densities using Monte Carlo simulations to compare the values retrieved from forward-modeled data with Poissonian noise and known input values. However, unlike in \citet{keeney17}, we found no evidence for systematic offsets. Reassuringly, the large optical depth of the \ce{H2O} absorption makes the fitting results more robust. Thus, we adopt the \ce{H2O} and \ce{O2} column densities returned by our fits as final.

\section{Discussion}
\label{sec:disc}

\begin{figure}
  \epsscale{1.15}
  \centering\plotone{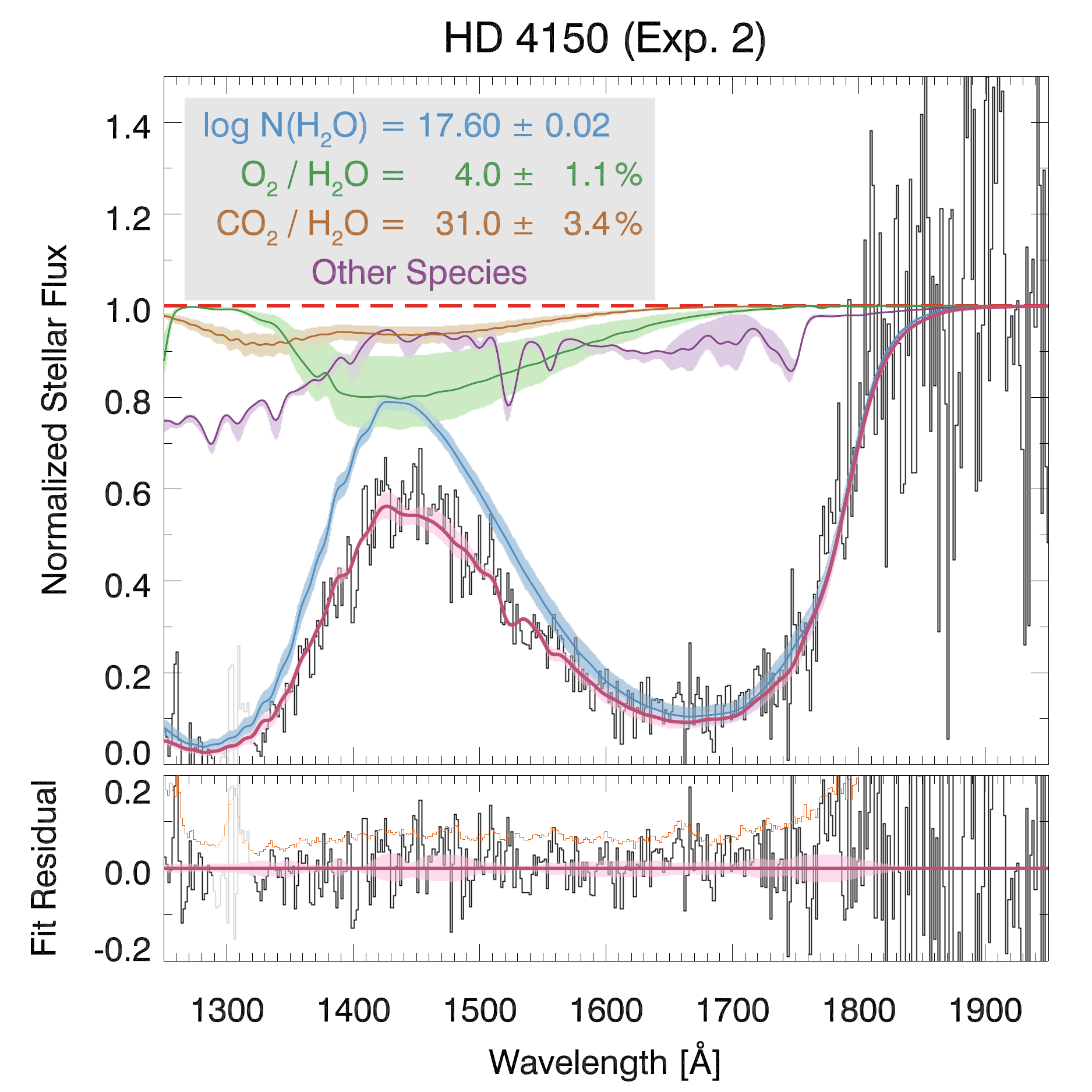}
  \vspace{-1ex}
  \caption{Best-fit column densities for the second post-occultation exposure of HD~4150, with 95\% ($2\sigma$) confidence bands.
  \label{fig:fit2}}
\end{figure}

The decrease in the \ce{H2O} column density from the first exposure of HD~4150 to the second (see \autoref{fig:fit1}-\ref{fig:fit2}) agrees with the na\"ive expectation that $N\propto\rho^{-1}$ \citep{haser57}, where $\rho$ is the impact parameter with respect to the nucleus. $N_{\ce{H2O}}$ decreases by a factor of $\sim1.4$ between the two exposures when $\rho$ increases by a factor of $\sim1.3$ (we estimate $\rho\approx0.3$~km and 0.4~km for the first and second exposures, respectively; see \autoref{fig:navcam}). However, the contemporaneous VIRTIS measurement \citep[$\log{N_{\ce{H2O}}}=17.00\pm0.04$ at $\rho\approx2.7$~km;][]{bockelee-morvan16}\footnote{\citet{bockelee-morvan16} list $\rho=3.6$~km for the contemporaneous exposure, which is measured from the center of the nucleus. The value we list is measured with respect to the limb of the nucleus for consistency with \autoref{fig:navcam}.} is not as consistent with $N\propto\rho^{-1}$, finding a factor of $\sim6$ decrease in $N_{\ce{H2O}}$ compared to the first post-occultation exposure when $\rho$ increases by a factor of $\sim9$.

Of course, the \citet{haser57} model, which assumes spherically symmetric outflow from a point source, is far too simplistic to be directly applied to data from \rosetta\ 67P/C-G. From \rosetta's vantage point embedded in the coma of 67P/C-G, the nucleus is clearly resolved and far from spherical (\autoref{fig:navcam}), and observed gas \citep[e.g.,][]{migliorini16} and dust \citep[e.g.,][]{gerig18} outbursts are not axisymmetric. Thus, it is questionable whether the analytic prediction of \citet[namely, $N\propto\rho^{-1}$]{haser57} holds for Alice data.

The best-fit values of the \ce{O2} column density are $\log{N_{\ce{O2}}}=16.82\pm0.04$ for the first post-occultation exposure (\autoref{fig:fit1}) and $\log{N_{\ce{O2}}}=16.20\pm0.12$ for the second exposure (\autoref{fig:fit2}), corresponding to a factor of $\sim4$ decrease. It is unclear why $N_{\ce{O2}}$ is decreasing almost three times more quickly than $N_{\ce{H2O}}$. \citet{keeney17} found no clear trend in $N_{\ce{O2}}$ as a function of impact parameter when $\rho\approx5$-15~km (see their Figure~9). However, the exposures of HD~4150 probe the coma much closer to the nucleus than any of our previous sight lines, so we cannot confidently extrapolate the \citet{keeney17} measurements to $\rho<1$~km.

The most puzzling result of \citet{keeney17} was that the relative abundance of \ce{O2} with respect to \ce{H2O} inferred from Alice data (median $N_{\ce{O2}}/N_{\ce{H2O}}=25$\%) was nearly an order of magnitude larger than the average value of $n_{\ce{O2}}/n_{\ce{H2O}}=3.85\pm0.85$\% found by ROSINA \citep{bieler15}. In fact, the datasets almost seemed to be mutually exclusive; \citet{keeney17} did not find a single high-quality example with $N_{\ce{O2}}/N_{\ce{H2O}}<10$\% (although there were several upper limits), and ROSINA almost never measured $n_{\ce{O2}}/n_{\ce{H2O}}>10$\% \citep{bieler15,fougere16}. \citet{keeney17} speculated that the discrepancy could be attributed to comparing column density along a line of sight near the nucleus with number density measured \textit{in situ} at the spacecraft position, but no firm conclusions could be drawn. It is therefore reassuring that we measure a relative abundance of $\ce{O2}/\ce{H2O}$ that is nearly identical to the average ROSINA value in one of the post-occultation exposures (see \autoref{fig:fit2}).

However, whereas \citet{keeney17} did not find any instances of $\ce{O2}/\ce{H2O}$ consistent with ROSINA measurements in absorption, the emission-line technique of \citet{feldman16} frequently infers $\ce{O2}/\ce{H2O}$ values near the sunward limb of the nucleus in Alice data that are consistent with ROSINA values. For example, an estimate of $\ce{O2}/\ce{H2O}$ can be derived from exposures taken just before and just after the occultation using the ratio of the semi-forbidden \ion{O}{1}] 1356~\AA\ line to \ion{H}{1} Ly$\beta$. If we assume that all of the \ion{O}{1}] 1356~\AA\ emission near the limb comes from electron impact on \ce{H2O} \citep{makarov04}, \ce{O2} \citep{kanik03}, and \ce{CO2} \citep{mumma72} at an energy of 100~eV, then the brightness of the \ion{C}{1} 1657~\AA\ line suggests that 25-30\% of the 1356~\AA\ brightness comes from electron impact dissociation of \ce{CO2}. Similarly, large off-nadir steps along the slit suggest that 25-30\% of the \ion{H}{1} Ly$\beta$ brightness comes from resonant scattering in the coma. After these corrections, we estimate that $\ce{O2}/\ce{H2O}\approx4$\% in the Alice exposures surrounding the occultation. Although this estimate is predicated on many assumptions, it is nevertheless reassuring that it is identical to the value we derived for our second exposure (see \autoref{fig:fit2}).

\edit1{Further, the above estimate is consistent with $\ce{O2}/\ce{H2O}\approx4$\% being the quiescent value at this time, whereas the elevated value of $\ce{O2}/\ce{H2O}\approx10$\% (\autoref{fig:fit1}) may be associated with a strong, collimated dust outburst observed in the sunward direction by VIRTIS-H \citep{bockelee-morvan17} and VIRTIS-M \citep{rinaldi18}.} The dust outburst peaked at approximately 13:30:00 UT \citep{rinaldi18}, $\sim20$~minutes before our first post-occultation exposure began. At the beginning of the first Alice exposure, the radiance of the dust emission had decayed to $\sim20$\% of its peak compared to the quiescent level, and by the beginning of the second Alice exposure it had returned to the quiescent level altogether \citep[see Figure~4 of][]{rinaldi18}. Thus, one plausible explanation for the differing $\ce{O2}/\ce{H2O}$ levels in the two Alice exposures is a non-constant production rate of \ce{O2} (i.e., the amount of \ce{O2} in the first exposure is affected by the dust outburst). \edit1{However, \citet{bockelee-morvan17} found no increase in \ce{H2O} or \ce{CO2} column density during the outburst, so it is unclear why the \ce{O2} production rate would be affected but not those of \ce{H2O} and \ce{CO2}.}

\begin{figure}
  \epsscale{1.15}
  \centering\plotone{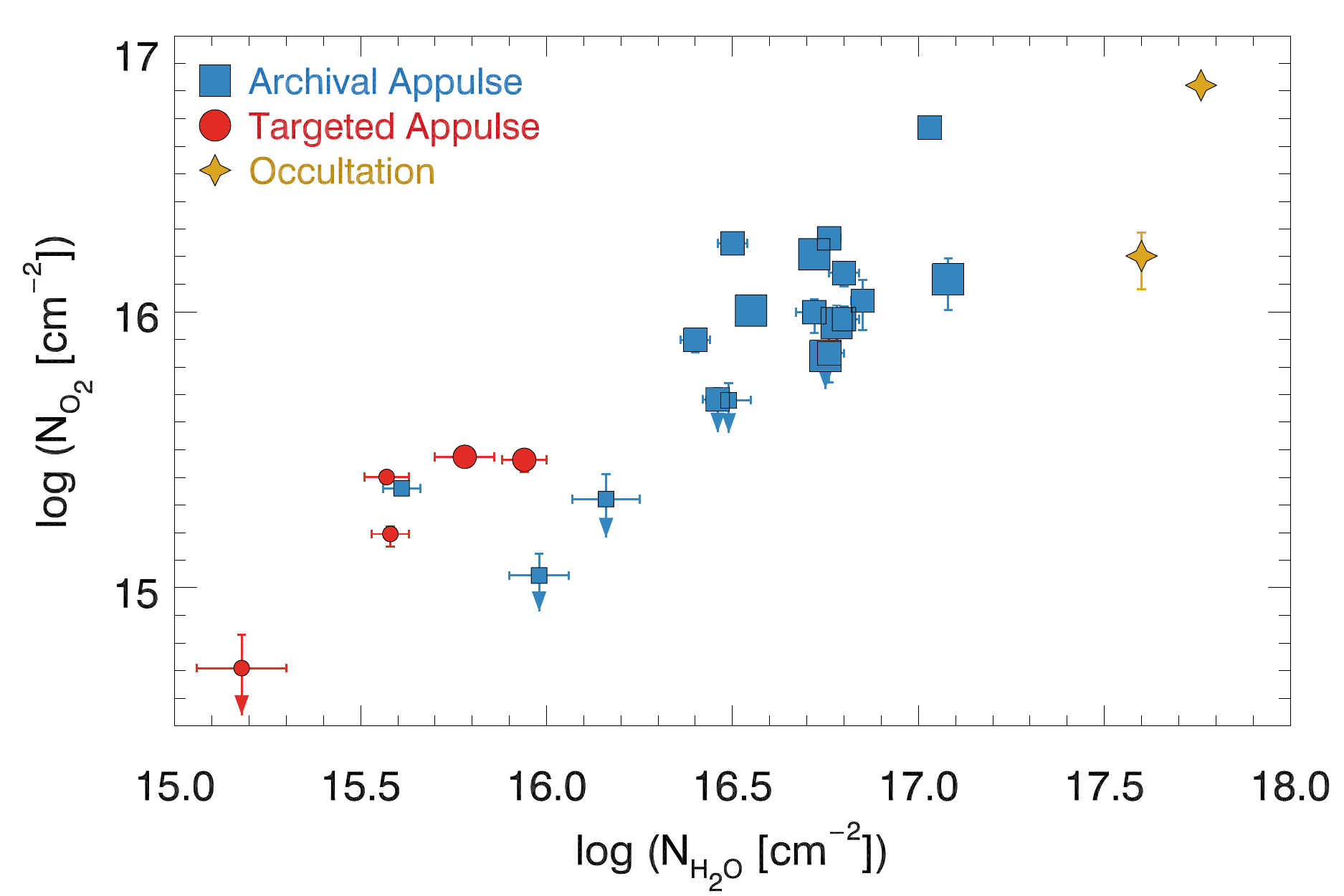}
  \centering\plotone{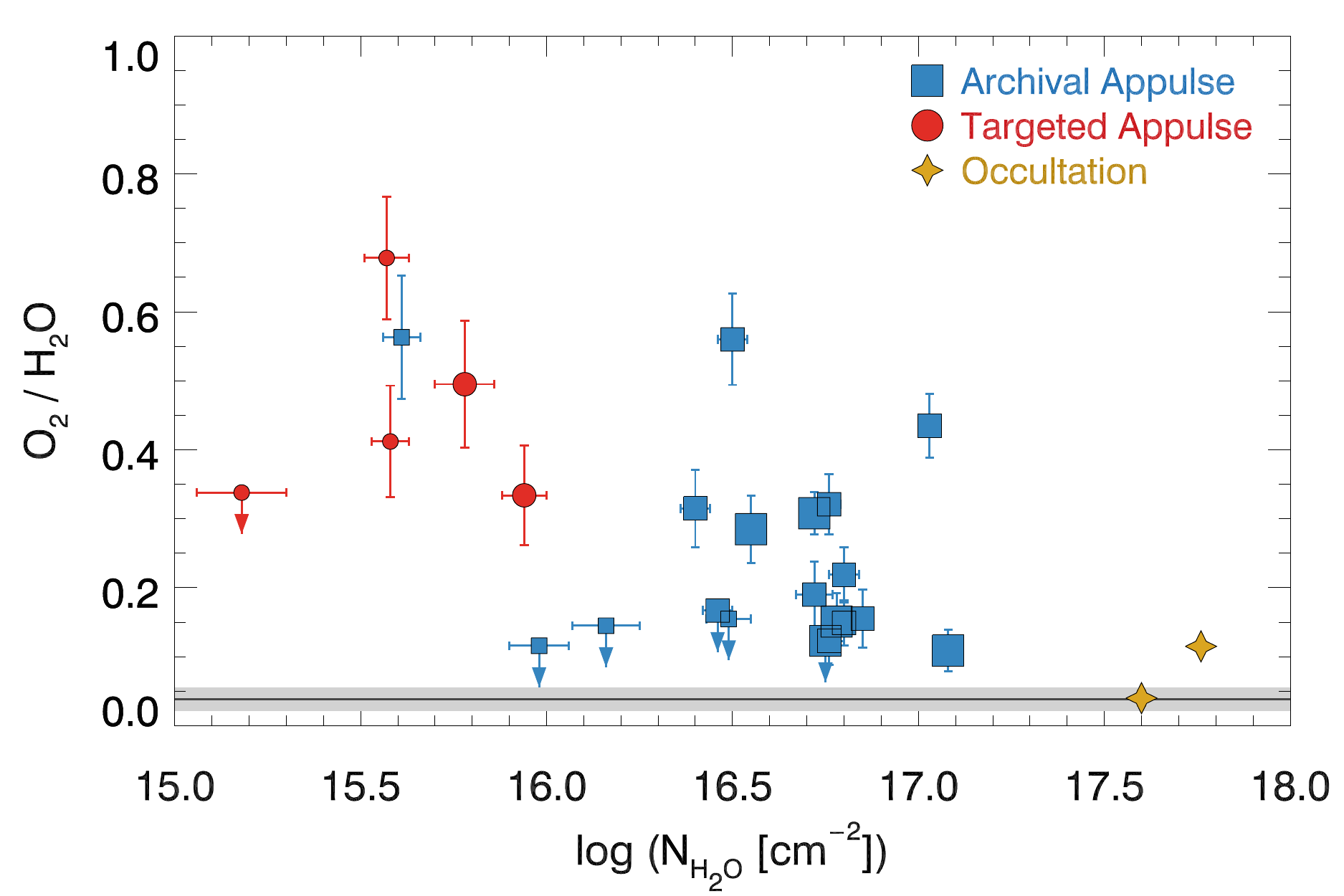}
  \vspace{-1ex}
  \caption{\ce{O2} column density (\textit{top}) and relative abundance of $\ce{O2}/\ce{H2O}$ (\textit{bottom}) as a function of $N_{\ce{H2O}}$ for the two post-occultation exposures of HD~4150 and the stellar appulses of \citet{keeney17}. There are clear trends of increasing $N_{\ce{O2}}$ and decreasing $\ce{O2}/\ce{H2O}$ with increasing $N_{\ce{H2O}}$. The shaded area in the bottom panel is the 95\% ($2\sigma$) confidence band for $n_{\ce{O2}}/n_{\ce{H2O}}$ from \citet{bieler15}.
  \label{fig:NH2O_O2}}
\end{figure}

\autoref{fig:NH2O_O2} shows $N_{\ce{O2}}$ and $N_{\ce{O2}}/N_{\ce{H2O}}$ as a function of $N_{\ce{H2O}}$ for the two post-occultation exposures of HD~4150 and the stellar appulses of \citet{keeney17}. \edit1{We confirm that $N_{\ce{O2}}$ and $N_{\ce{H2O}}$ are strongly correlated as expected due to the strong correlation between $n_{\ce{O2}}$ and $n_{\ce{H2O}}$ in ROSINA data \citep{bieler15,fougere16}.} Owing to the smaller impact parameters probed, our occultation data find larger $N_{\ce{H2O}}$ values and smaller $N_{\ce{O2}}/N_{\ce{H2O}}$ than the appulses, and overall there is a clear trend of decreasing $\ce{O2}/\ce{H2O}$ with increasing $N_{\ce{H2O}}$ in Alice data (e.g., when $N_{\ce{H2O}}<10^{16}~\mathrm{cm}^{-2}$ the median $N_{\ce{O2}}/N_{\ce{H2O}}=41$\%, whereas the median $N_{\ce{O2}}/N_{\ce{H2O}}=16$\% for larger $N_{\ce{H2O}}$).

This trend, \edit1{which was first noted by \citet{bieler15}}, suggests that we are able to detect an $\ce{O2}/\ce{H2O}$ abundance consistent with the ROSINA measurements in \autoref{fig:fit2} simply because the \ce{H2O} column density is sufficiently large. A $\mathrm{S/N}$-dependent detection threshold for $N_{\ce{O2}}$ in Alice data is consistent with this trend, but we see no clear evidence that our measurements are strongly affected by such a selection bias (\autoref{fig:NH2O_O2}). Further, this hypothesis cannot explain \textbf{all} of the Alice measurements since \citet{keeney17} found two examples where $N_{\ce{O2}}/N_{\ce{H2O}}>40$\% when $N_{\ce{H2O}}>10^{16}~\mathrm{cm}^{-2}$. Thus, a full explanation for the discrepancy in $\ce{O2}/\ce{H2O}$ abundance between Alice and ROSINA measurements remains elusive.

\edit1{\subsection{Empirical Coma Model Comparisons}}
\label{sec:disc:rosina}

\edit1{\citet{hansen16} developed an empirical coma model to study the evolution of the \ce{H2O} production rate between 2014~June and 2016~May. This model is based on comparisons between ROSINA data and Direct Simulation Monte Carlo models of the 3D neutral gas coma, and corroborated by comparisons with other \rosetta\ instruments (VIRTIS, MIRO, RPC) and ground-based dust measurements. Here we compare the predictions of the \citet{hansen16} model with \ce{H2O} column densities measured by Alice from stellar appulses and occultations.}

\edit1{\citet{hansen16} parameterize the number density, $n$, of \ce{H2O} molecules as}

\begin{equation}
  \label{eqn:density}
  n = \frac{f Q}{4\pi r^2 v}
\end{equation}

\edit1{\noindent where $Q$ is the \ce{H2O} production rate, $r$ is the distance from the comet center, $v$ is the gas velocity, and $f$ is an empirical correction factor. When $f=1$, \autoref{eqn:density} is equivalent to spherically symmetric radial expansion; otherwise, it accounts for the observed anisotropy in the coma of 67P/C-G \citep{fougere16,hansen16,lauter19}. Between the equinoxes (2015~May to 2016~March) the factors $Q$, $v$, and $f$ are independent of $r$ \citep[see Tables~1 and 2 of][]{hansen16}, so the number density can be integrated to find the column density:}

\begin{align}
  \label{eqn:coldens}
  N &= \frac{f Q}{4\pi v} \int_0^\infty \frac{ds}{r^2(s)} \\
  \nonumber
  &= \frac{f Q}{4\pi v} \int_0^\infty \frac{ds}{s^2 + r_{\rm sc}^2 - 2 s r_{\rm sc} \cos{\theta}}
\end{align}

\edit1{\noindent where $r_{\rm sc}$ is the distance from the spacecraft to the comet center, $\theta$ is the angle of the sight line with respect to the comet center, and $r^2(s)$ is given by the law of cosines. The integration is performed along the line of sight, $s$, starting from the spacecraft position ($s\equiv0$).}

\edit1{ Prior to the inbound equinox, $f$ is a function of both $R_h$ and $r$ \citep{hansen16} and cannot be separated from the integral. $Q$ and $v$ are defined in Equations 7 and 10 of \citet{hansen16}, and depend only on heliocentric distance:}

\begin{align}
  \nonumber
  Q(R_h) &= 
  \begin{cases}
    (2.58\pm0.12)\,R_h^{-5.10\pm0.05}, & \text{pre-perihelion} \\
    (15.8\pm0.9)\,R_h^{-7.15\pm0.08},  & \text{post-perihelion}
  \end{cases} \\
  \nonumber
  v(R_h) &= (771.0 - 55.5\,R_h)\left(1 + 0.171\,e^{-\frac{R_h-1.24}{0.13}}\right)
\end{align}

\edit1{\noindent where $Q$ has units of $10^{28}~\mathrm{molecules\,s^{-1}}$ and $v$ has units of $\mathrm{m\,s^{-1}}$.}

\begin{figure}
  \epsscale{1.15}
  \centering\plotone{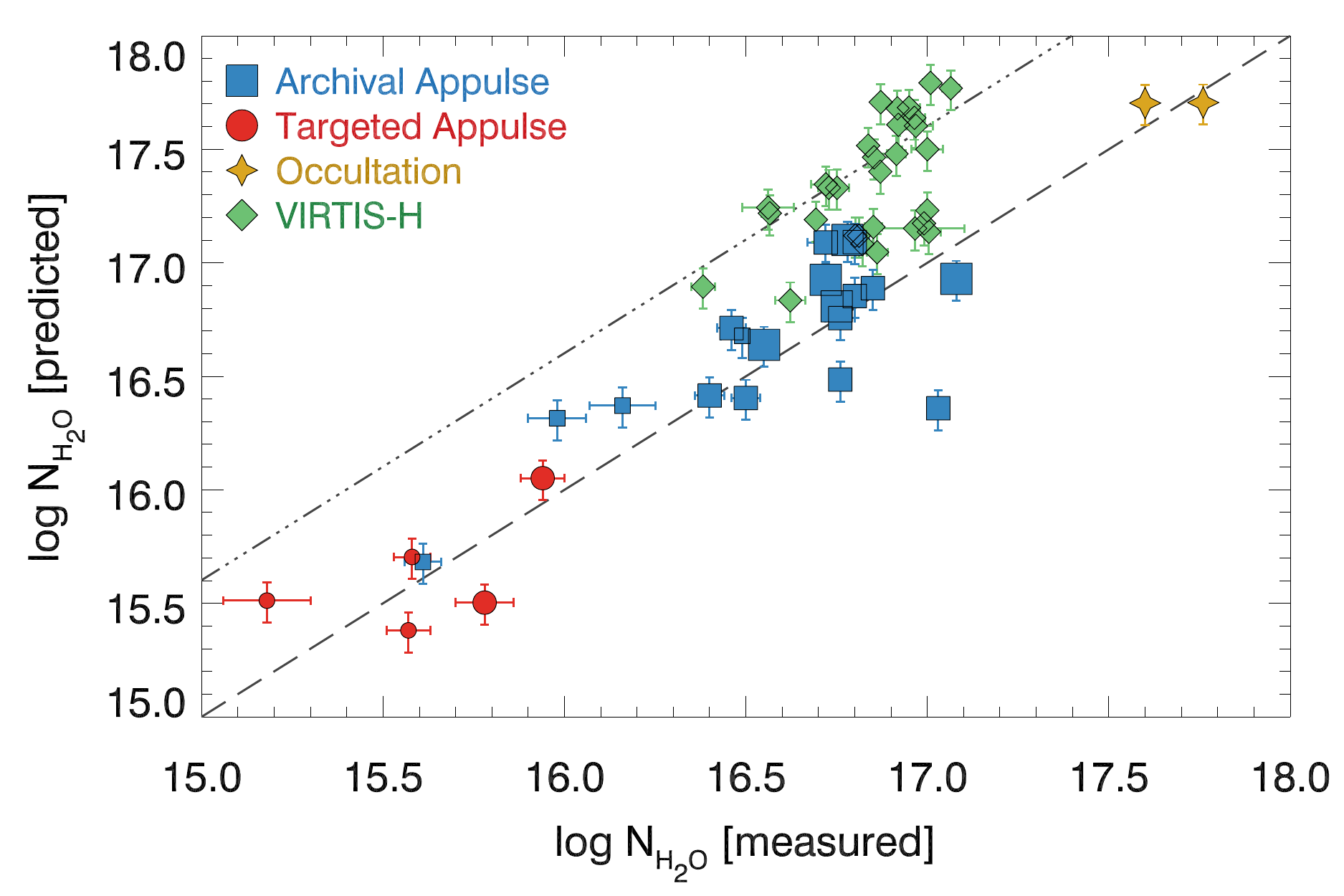}
  \vspace{-1ex}
  \caption{\edit1{The predicted \ce{H2O} column density from \autoref{eqn:coldens} compared to Alice and VIRTIS measurements. The dashed line shows perfect agreement between the predicted and measured values, and the dot-dashed line shows predicted values that are four times larger than measured.}
  \label{fig:NH2O_hansen}}
\end{figure}

\begin{figure}
  \epsscale{1.15}
  \centering\plotone{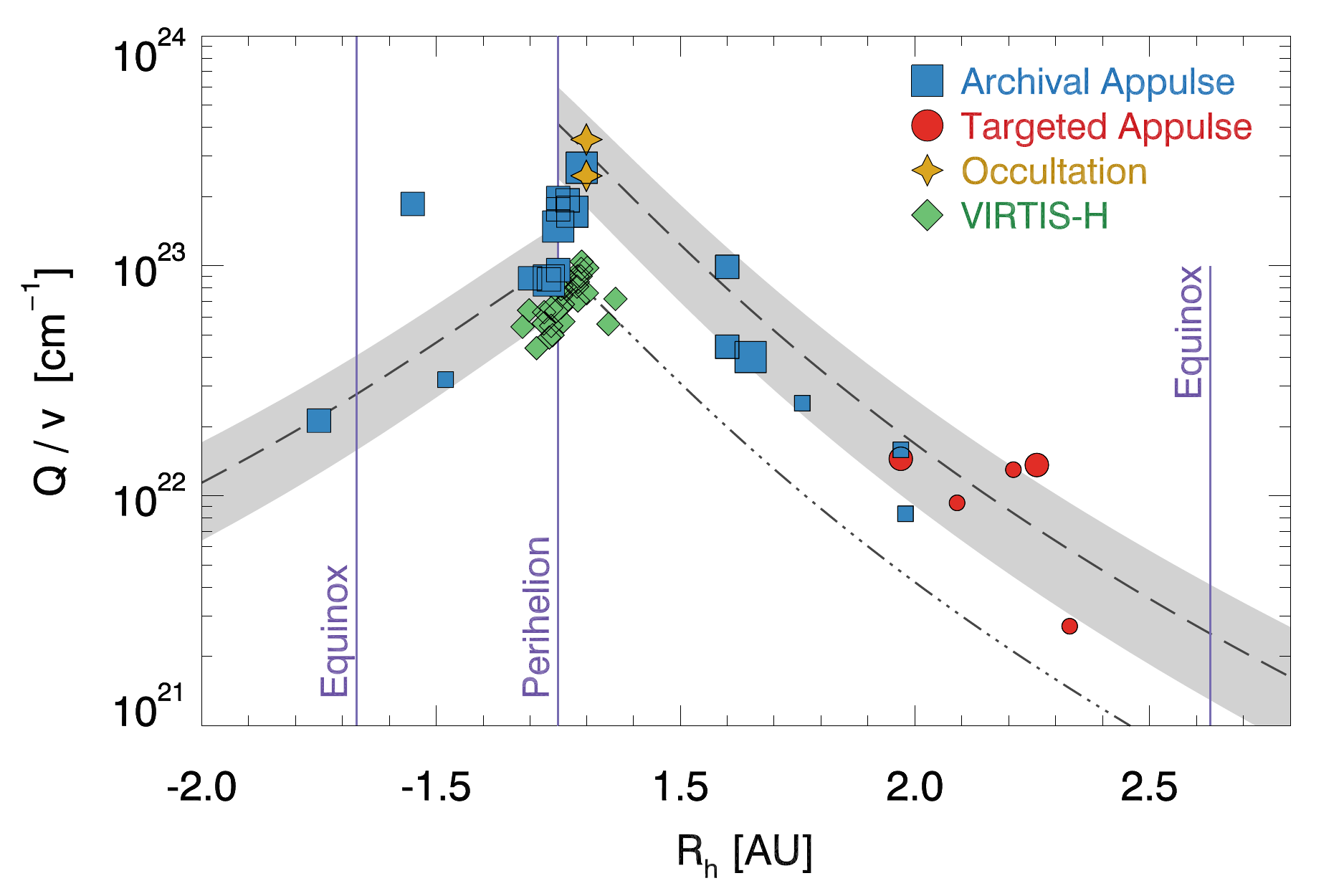}
  \vspace{-1ex}
  \caption{\edit1{The predicted trend of $Q/v$ as a function of $R_h$ \citep{hansen16}, compared to values inferred from Alice and VIRTIS measurements using \autoref{eqn:trend}. The shaded region shows the 95\% ($2\sigma$) confidence band for $Q/v$ and the dashed lines show the best-fit relation from \citet{hansen16}; the dot-dashed line post-perihelion shows the \citet{hansen16} prediction reduced by a factor of four.}
  \label{fig:Rh_hansen}}
\end{figure}

\edit1{\autoref{fig:NH2O_hansen} shows the predicted \ce{H2O} column density from \autoref{eqn:coldens} compared to Alice and VIRTIS measurements near perihelion. The uncertainty on the model predictions is assumed to be 20\% \citep{hansen16}. The Alice data are consistent with the predicted values over the full range of column densities measured. Using all data obtained between the equinoxes (all but one data point, see \autoref{fig:Rh_hansen}), where the assumptions behind \autoref{eqn:coldens} are valid, the RMS difference between the Alice measurements and the \citet{hansen16} prediction is 0.24~dex.}

\edit1{However, for VIRTIS data \citep{bockelee-morvan16}, the predicted column densities are systematically higher than the measurements. \citet{fougere16} had to reduce the \ce{H2O} column densities predicted by their model by a factor of four to match the measurements of \citet{bockelee-morvan16}. The dot-dashed line in \autoref{fig:NH2O_hansen} indicates a model over-prediction by a factor of four and passes through much of the VIRTIS data, agreeing with the analysis of \citet{fougere16}.}

\edit1{As a final check on the consistency of the Alice data and the \citet{hansen16} coma model, \autoref{eqn:coldens} can be rearranged so that the details of the observing geometry cancel out, leaving only a predicted trend with heliocentric distance:}

\begin{equation}
  \label{eqn:trend}
  \frac{Q(R_h)}{v(R_h)} = \frac{4\pi N/f}{\int_0^\infty (s^2 + r_{\rm sc}^2 - 2 s r_{\rm sc} \cos{\theta})^{-1} ds}
\end{equation}

\edit1{\noindent \autoref{fig:Rh_hansen} shows the \citet{hansen16} prediction of $Q/v$ as a function of $R_h$ compared to values derived from Alice and VIRTIS measurements using \autoref{eqn:trend}. The dashed lines show the best-fit relation of \citet{hansen16}, and the gray shaded regions show the 95\% ($2\sigma$) confidence band for the prediction. The Alice appulse \citep{keeney17} and occultation measurements (\autoref{fig:fit1}-\ref{fig:fit2}) are consistent with the \citet{hansen16} prediction, except very close to perihelion where the \citet{hansen16} model is discontinuous.}

\edit1{The VIRTIS-H data are lower than the \citet{hansen16} prediction over the full range of heliocentric distance, and show no discontinuity at perihelion. The disagreement is largest after perihelion, where the VIRTIS data are lower than the prediction by a factor of $\sim4$ (dot-dashed line). \citet{fougere16} discussed possible reasons for this discrepancy and we will not speculate further, except to note that it is reassuring that we are able to reproduce the reported tension.}

\edit1{Furthermore, the fact that our \ce{H2O} measurements are consistent with the predictions of the \citet{hansen16} coma model increases confidence in our ability to directly compare results from Alice and ROSINA. On the other hand, the broad consistency in \ce{H2O} results between the two instruments makes the discrepancy in $\ce{O2}/\ce{H2O}$ values all the more puzzling.}

\edit1{\subsection{Modeling Uncertainties}}
\label{sec:disc:modeling}

\edit1{There are several sources of systematic uncertainty in our modeling process, as detailed in \citet{keeney17}. Perhaps the hardest to quantify is the uncertainty introduced by unknown far-UV absorption cross sections. ROSINA-DFMS has found dozens of species in the coma of 67P/C-G \citep{leroy15,altwegg17}, many of which have no measured far-UV cross sections, and could thus conceivably have a large enough cross section that even a trace amount could produce appreciable absorption. We believe this circumstance is unlikely but cannot rule it out.}

\edit1{Another concern is temperature dependence of the adopted cross sections, because all of the cross sections we use were measured at 250-300~K \citep[see Table~3 of][]{keeney17},} \edit2{and the coma is expected to cool adiabatically.} \edit1{There are several reasons for this choice. First, high-resolution laboratory measurements are not consistently available for all modeled species at any other temperature. Adopting a consistent temperature for the cross sections of all modeled species minimizes the uncertainty introduced by cross sections for different species having different behavior with decreasing temperature.}


\edit1{Reassuringly, \autoref{fig:NH2O_hansen} shows that Alice measurements of \ce{H2O}, derived from absorption cross sections measured at 250~K \citep{chung01}, are consistent with the predictions of the \citet{hansen16} empirical coma model for 67P/C-G. Thus, if temperature-dependent cross sections are invoked to explain different $\ce{O2}/\ce{H2O}$ measurements from ROSINA and Alice, they must preferentially affect \ce{O2} or else the Alice and ROSINA \ce{H2O} values would disagree.}

\edit1{Finally, and most speculatively, there are \ce{O2} and \ce{H2O} cross-section measurements at multiple temperatures over part of the modeled far-UV wavelength range which suggest that using room-temperature cross sections yields lower $\ce{O2}/\ce{H2O}$ values than at lower temperatures. Low-resolution measurements \citep{yoshino05} indicate that the peak \ce{O2} cross-section at $\sim1400$~\AA\ decreases by $\sim0.1$~dex as the temperature decreases from 295 to 78~K, implying that a larger \ce{O2} column density is required to match the observed absorption at lower temperatures. High-resolution \ce{H2O} cross sections are available at 250~K \citep{chung01} and 298~K \citep{mota05}, and the peak cross section at $\sim1650$~\AA\ is $\sim0.1$~dex larger at 250~K, implying that a smaller \ce{H2O} column density is required to match the observed absorption at lower temperatures. Although we cannot be confident that the \ce{H2O} cross section continues to increase at temperatures $<250$~K, the existing data indicate that lower temperatures necessitate larger $\ce{O2}/\ce{H2O}$ to fit the data, further suggesting that adopting room-temperature cross sections is not the cause of the discrepancy between our $\ce{O2}/\ce{H2O}$ values and the ROSINA measurements.}

\section{Summary}
\label{sec:summary}

We have presented far-UV spectra of the A0~IV star HD~4150, which was serendipitously observed by \rosetta's Alice imaging spectrometer on 2015 September~13. HD~4150 was observed in two 10-minute integrations immediately after being occulted by 67P/C-G, and revisited approximately 9~months later when its line of sight was far from the nucleus. By comparing the two epochs of stellar spectra, we were able to quantify the amount of \ce{H2O} and \ce{O2} within 1~km of the nucleus.

We find that $N_{\ce{H2O}}\propto\rho^{-1}$ in our consecutive exposures of HD~4150, but $N_{\ce{O2}}$ decreases $\sim3$ times faster than $N_{\ce{H2O}}$. We have also measured a value of $N_{\ce{O2}}/N_{\ce{H2O}}$ that is consistent with ROSINA measurements of $n_{\ce{O2}}/n_{\ce{H2O}}$ \citep{bieler15,fougere16}. \edit1{Combining our observations of HD~4150 with previous results from \citet{keeney17}}, we confirm the strong correlation between \ce{O2} and \ce{H2O} \edit2{(see top panel of \autoref{fig:NH2O_O2}, which demonstrates that $N_{\ce{O2}}$ increases as $N_{\ce{H2O}}$ increases)} first reported by ROSINA \citep{bieler15,fougere16}, \edit2{but} find a general decrease in $\ce{O2}/\ce{H2O}$ with increasing $N_{\ce{H2O}}$ in Alice data \edit2{(see bottom panel of \autoref{fig:NH2O_O2})}\edit1{, even though the HD~4150 data in isolation suggest otherwise}. This trend of decreasing $\ce{O2}/\ce{H2O}$ with increasing $N_{\ce{H2O}}$ partially explains the initial discrepancy in $\ce{O2}/\ce{H2O}$ between Alice and ROSINA.

Several \rosetta\ instruments (e.g., Alice, ROSINA, VIRTIS, MIRO) can detect \ce{H2O} in the coma of 67P/C-G. However, only Alice and ROSINA can directly detect \ce{O2}, so the differing Alice and ROSINA measurements of $\ce{O2}/\ce{H2O}$ in the coma of 67P/C-G remain mysterious, \edit1{especially since we have shown that our \ce{H2O} measurements are consistent with the empirical coma model of \citet{hansen16}.}

\edit1{We have investigated several potential sources of this discrepancy, but none have provided a satisfactory explanation.} \edit2{We note, however, that additional high-resolution laboratory measurements of molecular absorption cross sections in the far-UV at temperatures of $\sim100$~K would be welcome.}

\edit1{One avenue of future study is modeling the $\ce{O2}/\ce{H2O}$ distribution throughout the coma of 67P/C-G, then integrating along the Alice line of sight to predict the observed column density ratio. The models of \citet{hansen16}, \citet{fougere16}, and \citet{lauter19} are based off of \textit{in-situ} ROSINA samples of the coma density and composition, and largely agree with emission-line measurements from MIRO \citep{biver15,lee15}, VIRTIS \citep{bockelee-morvan15,fink16}, and Alice \citep{feldman16}. However, emission-line intensities are far more sensitive to density than absorption ($n^2$ compared to $n$ dependence), so a distributed source of \ce{O2} would preferentially reveal itself in absorption.}

\acknowledgments
\rosetta\ is an ESA mission with contributions from its member states and NASA. We thank the members of the \rosetta\ Science Ground System and Mission Operations Center teams, in particular Richard Moissl and Michael K\"uppers, for their expert and dedicated help in planning and executing the Alice observations. The Alice team acknowledges continuing support from NASA via Jet Propulsion Laboratory contract 1336850 to the Southwest Research Institute. This research has made use of the SIMBAD database, operated at CDS, Strasbourg, France.

\facility{\rosetta\ (Alice)}

\bibliographystyle{aasjournal}
\bibliography{references}

\end{document}